\documentclass[]{article}
\usepackage{amsmath,amsfonts,a4wide}
\usepackage{graphicx,nicefrac}
\usepackage{color,subfigure}
\usepackage{authblk}

\newcommand{\diag}{\mathrm{diag}}
\newcommand{\by}{\times}

\title{Non-linearity in monochromatic transmission tomography}
\author[1]{William R.B. Lionheart}
\author[2]{Bj{\o}rn Tore Hjertaker} 
\author[2]{Rachid Maad} 
\author[3]{Ilker Meric} 
\author[4]{ Sophia B. Coban}
\author[5]{ Geir Anton Johansen} 

\affil[1]{School of Mathematics, University of Manchester}
\affil[2]{Department of Physics and Technology, University of Bergen}
\affil[3]{Department of Electrical Engineering,  Western Norway University of Applied Sciences}
\affil[4]{Centrum Wiskunde \& Informatica, Amsterdam}
\affil[5]{Faculty of Engineering and Business Administration,  Western Norway University of Applied Sciencese}
\date{}

\begin{document}

\maketitle

\begin{abstract}
While it is well known that X-ray tomography using a polychromatic source is non-linear, as the linear attenuation coefficient depends on the wavelength of the X-rays, tomography using near monochromatic sources are usually assumed to be a linear inverse problem. When sources and detectors are not treated as points the measurements are the integrals of the exponentials of line integrals and hence non-linear. In this paper we show that this non-linearity can be observed in realistic situations using both experimental measurements in a $\gamma$-ray tomography system and 
simulations. 
We exhibit the Jacobian matrix of the non-linear forward problem. We also demonstrate a reconstruction algorithm, which we apply  to experimental data and we show that improved reconstructions can be obtained over the linear approximation. 
\end{abstract}

\section{Introduction}

X-ray or $\gamma$-ray transmission tomography (CT) is generally taken to be a linear method in that the vector of the logarithm of the measurements on each detector, for each source, is a linear function of the linear attenuation in the region being imaged. Under this assumption the problem of recovering the linear attenuation in each pixel (or voxel) in the image reduces to one of linear algebra. The matrix of the resulting linear system will be mildly ill-conditioned (at least for fairly complete data) and standard regularized inversion methods or discrete approximations to explicit reconstruction algorithms for the continuum case work reasonably well \cite{natterer1986mathematics}.  There are several reasons for a linear model not to fit CT data. In the polychromatic case the dependence of the linear attenuation on wavelength is often the largest effect and is important in medical, dental, non-destructive testing and in security screening applications. Scattering also results in non-linearity \cite{glover1980nonlinear}, as well as a forward problem that more highly coupled as material out of the direct beam path between a source and detector can affect the measurement. This has been identified as the \textit{penumbra} effect \cite{Kingston2015,Kueh2016}. This can be counteracted by collimation of the detectors using a \lq scatter grid' but in some systems such as the fast switched sourced RTT security screening CT system this is not possible \cite{thompson2015high, wadeson2010scatter}. However even nearly monochromatic CT systems using for example a $\gamma$-ray or synchrotron source, and even without scatter, can exhibit non-linearity.

For simplicity we will consider a two dimensional problem (generalization to three dimensions is straightforward).  Consider a fixed source and detector pair both represented by a line segment. Let $S(s)$ be a point on the source point parametrized by arc length $s$ and $D(s)$ similarly a point on the detector. From the Beer-Lambert law \cite{BeerA}, flux density $I$ measured at a point in the detector due to a point in the source with flux density $I_0$ is 
\begin{equation}
I(D(s))= I_0\left(S(s')\right)\exp\left( -\int\limits_{L\left(S(s'),D(s)\right)}\mu\,\mathrm{d}\ell\right),
\end{equation}
where the integral is taken over the line $L\left(S(s'),D(s)\right)$ between source and detector points, and $\mathrm{d}\ell$ is the measure on that line.  The total flux for this source-detector pair is thus 
\begin{equation}
I = \iint I_0\left(S(s')\right)\exp\left( -\int\limits_{L\left(S(s'),D(s)\right)}\mu\,\mathrm{d}\ell\right) \,\mathrm{d}s' \, \mathrm{d}s,
\end{equation}
and we see the problem: the sum of exponentials is not the exponential of the sum.  This non-linearity is clearly bigger for larger sources and detectors giving a wider range of ray paths, and also increases as the variation in $\mu$ is such that the ray paths in a source and detector pair encounter a wider range of linear attenuation. The non-linearity has been described as a \lq\lq non-linear partial volume effect\rq\rq \cite{glover1980nonlinear}, the volume here referring to the region defined by all the rays between a source and detector pair and the fact that part of that volume has a different attenuation from another part and the exponentials of the integrals are combined. It has also been described in the literature as ``exponential edge-gradient effect'' \cite{joseph1981exponential, de1999metal}. For essentially two dimensional problems the phenomena has been noticed in the out of plane direction when the attenuation varies on the length scale of the slice width. According to one review paper ``Hopefully, this effect on the variance is small because accounting for it seems to be challenging'' \cite{nuyts2013modelling}. 

When using large sources and detectors it has been noticed that the non uniform density of rays across the volume needs to be accounted for \cite{yu2012finite, Kingston2015} and that this can improve reconstruction under some circumstances even without taking account of the non-linearity.

\section{Mathematical formulation}
The starting point of this work is to consider a finite set of  positions of sources and detectors of finite size. Although we can consider the continuous case of the Radon or X-ray transform where there are infinitely many point sources and detectors, there is no obvious candidate for a limiting continuum case with infinitely many measurements with  sources and detectors of non-zero size. Our investigation will therefore concern finitely many measurements. We can only recover a finite number of parameters in the the image space so we
suppose for simplicity that the linear attenuation coefficient can be represented as 
\begin{equation}
\mu(x) = \sum\limits_{k=1}^{K^2} \mu_k \chi_k(x),
\end{equation}
where $\chi_k$ are the characteristic functions of some partition of the region of interest in to $K^2$ picture elements, such as square pixels.  We will denote by $\mathbf{m}$ the vector of non-negative coefficients $\mu_i$.  We will assume that measurements are taken on $K^2$ pairs of line segments $S_k D_k$ thought of as a source, detector pair. 

Our data vector $\mathbf{d}$ is given by the non-linear forward problem
\begin{equation}
\mathbf{d}= \mathbf{F}(\mathbf{m}),
\end{equation}
where 
\begin{equation}
F_k(\mathbf{m}) =  
\log\left( \iint I_0\left(S_k(s')\right)\exp\left( -\int\limits_{L_{\left(S_k(s'),D_k(s)\right)}}\mu\,\mathrm{d}\ell\right) \,\mathrm{d}s' \, \mathrm{d}s \right).
\end{equation}
The Jacobian matrix, ${\partial F_k}/{\partial\mu_l}$ is 
\begin{equation}
-\frac{1}{\exp{F_k(\mathbf{m})}} \iint I_0\left(S_k(s')\right)\exp\left( -\int\limits_{L_{\left(S_k(s'),D_k(s)\right)}}\mu\, \mathrm{d}\ell\right) \left( \int\limits_{L_{\left(S_k(s'),D_k(s)\right)}}\chi_l \right) \, \mathrm{d}s' \, \mathrm{d}s. 
\end{equation}
In the simple case $I_0=1$, the derivative at $\mathbf{m}=0$ is simply
\begin{equation}
\frac{\partial F_k(\mathbf{0})}{\partial \mu_l} = -\frac{1}{\exp{F_k(\mathbf{0})}}
  \iint   \int\limits_{L_{\left(S_k(s'),D_k(s)\right)}}\chi_l\, \mathrm{d}\ell \,\mathrm{d}s' \, \mathrm{d}s,
\end{equation}
which coincides with the usual linearized problem for CT taking account of non-zero source and detector size \cite{yu2012finite,Kingston2015}.

\begin{figure}\centering
\includegraphics[width=0.8\textwidth]{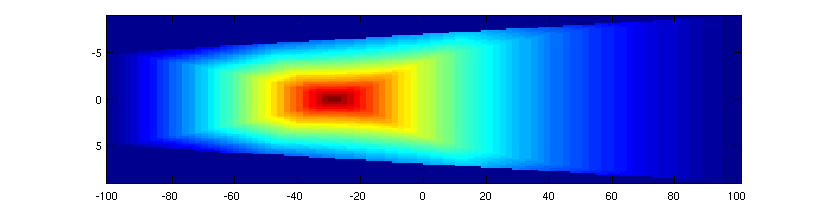}\\
\includegraphics[width=0.6\textwidth]{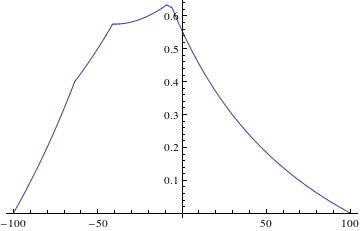}
\caption{The normalized linearized sensitivity for a simple case where source and detector are parallel and opposite. In this case the dimensions in mm are $W_1=18,W_2=10,L=200$. The lower plot shows a cross section for $y=3$ illustrating that the function is continuous but not differentiable.}
\label{fig:sensplot}
\end{figure}

In the limiting case, as the pixels tend to points, the integral reduces to the \textit{point sensitivity}, which is the density of rays at a point in the image. The point sensitivity for a simple case where source and detector are parallel and opposite with source width $W_1$, detector width $W_2$ separated by $L$ and $I_0=1$ is  
\begin{align}
S(x,y) = \frac{1}{2W_1 W_2}\,r_{1,2}\,r_{2,1},
\end{align}
where 
\begin{equation}\label{rij}
r_{i,j} = r\left(-W_i/2, W_i/2, y-\frac{d_j(L+2x)}{2(L-2x)}, y+\frac{d_j(L+2x)}{2(L-2x)}\right)
\end{equation}
and $r(a,b,c,d)$ is the length of the intersection of the intervals $(a,b)$ and $(c,d)$ (see Appendix A).
This rather complicated formula that simply expresses the product of the lengths of the line segments that are subsets of the source and detector that can be connected by straight lines to the point$(x,y)$ in question. While this is continuous the function is only piece-wise differentiable, and as can be seen from the example in fig.\ref{fig:sensplot} markedly inhomogeneous. Perhaps the most notable feature is that a `small' object is not detected close to the source or detector or at the edges of the `beam' joining the source and detector. Even for systems where the non-linearity is not especially pronounced this non-uniform sensitivity might still be sufficient to have a substantial effect on the accuracy of reconstructed images.

\begin{figure}\centering
\includegraphics[width=0.8\textwidth]{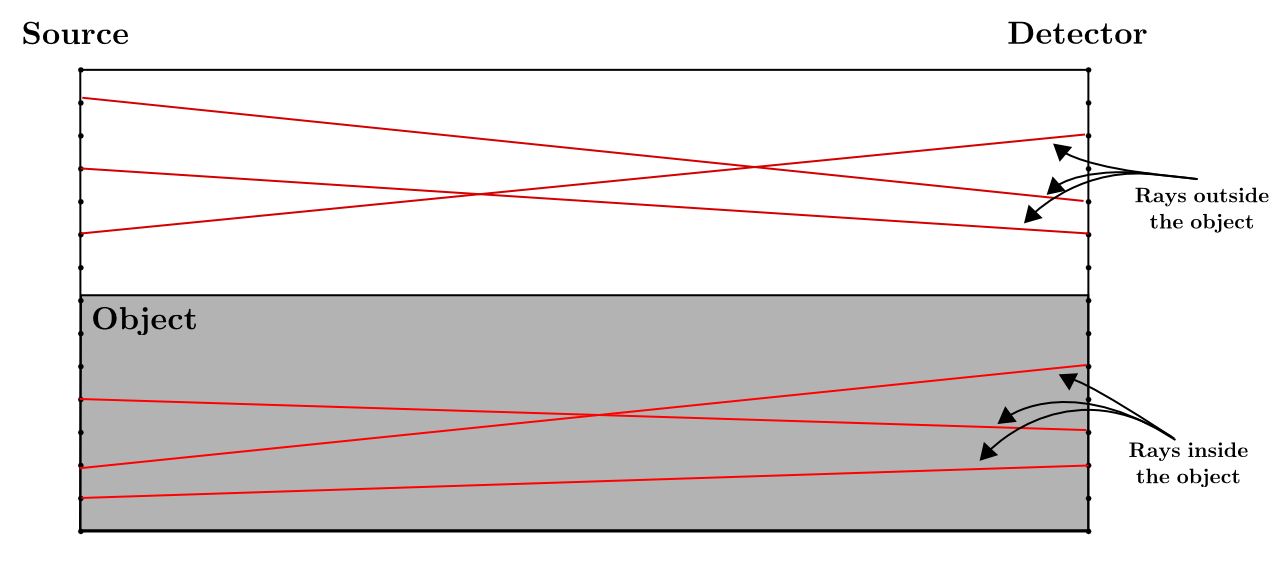}
\caption{A simple diagram illustrating NLPVE.}
\label{fig:nlpve_diag}
\end{figure}

The non-linear effect can be illustrated nicely with a source and detector parallel and the same width. First consider an object which occupies half width of the rectangle  between the source and detector. As a very crude approximation to illustrate the effect, consider only rays entirely inside the object (linear attenuation $\mu$) or entirely missing the object (linear attenuation $0$) and assume that the distance $L$ between source and detector is sufficiently larger than the width $W$ that we can ignore the variation in the path lengths. We have approximately the measurement as a function of $\mu$ 
$$F(\mu) = \log  \frac{1}{2}W \left(1+ \exp(-\mu L)  \right).$$
In this case we have, by taking series expansions,
$$ F(\mu)- F(0) =  -\frac{\mu L}{2}  +  \frac{\mu^2 L^2}{8} +O(\mu L)^3,$$ 
which gives us a `rule of thumb' that the magnitude of the  non-linear partial volume effect in this case is around half the square of the linear approximation to the difference in the logarithmic data. A more rigorous argument computing the integrals for all the rays between this source and detector and calculating the series with Mathematica agrees with this crude and intuitive argument. Splitting the rectangle the other way so that all the rays go half through the material is expected to produce a much smaller effect as it is only the variation in the path length through the materials that contributes to the exponentials of different numbers being summed before the logarithm is taken. In this case the series expansion for the difference in the logarithmic measurement due to taking account of the width $W$ is $ 1+ (W/L)^2+ O(W^3)$ showing that the effect can be considered as second order in the small parameter $W/L$. 

A simple approximation to the integral over lines joining source and detector can be achieved by choosing uniformly spaced quadrature points on each source and detector, and summing the intersection length with each pixel over all the lines joining quadrature points on each source and detector pair. Suppose there are $N$ pairs of quadrature points for each source-detector pair.  Let $R$ be the $NK\times K$ matrix of line intersection lengths of pixels of lines joining quadrature points. Let $P$ be the $K \times NK$ matrix that performs the (weighted) sum over the all the lines in a source detector pair. In the linear approximation to CT the matrix $PR$ would be the  matrix of the system we seek to solve.  Instead, for non-negative $\mathbf{d}$, we solve
\begin{equation}
\mathbf{d} = {F}(\mathbf{m}):=\log P \exp(R \mathbf{m}),
\end{equation}
where the $\log$ and $\exp$ applied to vectors are taken to act component-wise. The Fr\'echet derivative is
\begin{equation}
D {F}(\mathbf{m}) \delta \mathbf{m}= \left(\diag P \exp(R\mathbf{m})\right)^{-1} P\diag\left(\exp(R\mathbf{m} )\right) R \delta \mathbf{m}.
\end{equation}
Suppose that we have chosen pixels and source detector pairs such that $PR$ is invertible then by continuity the Fr\'echet derivative is also invertible for $\| \mathbf{m} \|$ sufficiently small. From the Inverse Function Theorem we see that ${F}$ is invertible on its range in a neighbourhood of $\mathbf{0}$. 
This local uniqueness result for the formally correctly determined problem simply reassures us that as far as uniqueness of solution is concerned the non-linear problem for small attenuation is no worse than the linear case. The general case remains an open problem.

\section{Reconstruction algorithms}\label{sec:recon_algs}


Let $p$ be the number of projections, $r$ be the number of rays per projection, $M = p\,r$, and $K$ be the length (and width) of the image domain in pixels. For both linear and non-linear reconstructions, the image $\bf{m}$ is reconstructed over a square grid of $K\by K$. To approximate the integrals over the source and detector we choose $N_S$ and $N_D$ quadrature points on the source and detector respectively and $N=N_D N_S$. We denote by $R$ the geometry matrix of size $N M\by K^2$. Each row of $R$ is obtained using Jacob's ray tracing algorithm \cite{JacobsRayTracing} to find the intersection length of a point between a quadrature point on the source and detector. The matrix $P$ is the $M \by NM $ matrix that simply takes a weighted sum of rows corresponding to each ray within a given source-detector pair. We denote by $\bf{d}$ the (vectorized) measured data with size $M\by 1$. We found by numerical experiment that for our geometry using numbers of quadrature points larger than $N_S=N_D=5$ did not change the results to three decimal places.

For the linear model, we need to solve $PR\mathbf{m} = \mathbf{d}$, although it will be overdetermined if we choose $K^2<M$ so we require a least squares solution
\begin{align*}
\arg\min_{\mathbf{m}} \|PR\mathbf{m}-\mathbf{d}\|^2
\end{align*}
using a straightforward implementation of conjugate gradient method adapted to least squares problems (henceforth CGLS) \cite{CGLSBjorck}. The CGLS method is essentially the conjugate gradient method applied to solve the least squares problem. 
CGLS is a popular method amongst those working in signal and image processing for its simple and computationally inexpensive implementation and fast convergence. For our reconstructions, the CGLS iterations were performed until a tolerance value, $\epsilon$, was reached or a maximum number of iterations, $i_{\mathrm{max}}$, was performed. The tolerance value is defined as the 2-norm of the residual vector calculated at each iteration. 
The algorithm was also applied to the generalized Tikhonov regularization problem \cite{Hansen}
\begin{align*}
\arg\min_{\mathbf{m}} \|PR\mathbf{m}-\mathbf{d}\|^2 + \alpha^2\|L\mathbf{m}\|^2,
\end{align*}
where $L$ is the regularization matrix and $\alpha>0$ 
is the regularization factor. For our runs for both numerically simulated and real $\gamma$-ray data reconstructions, the regularization matrix $L$ was chosen to be the identity matrix, and $\alpha = 0.01$. Another obvious choice of $L$ would be a finite difference gradient which would penalize less smooth images. In that case a larger $K$ could be used as the regularized system, formed by appending $L$ beneath $PR$, would still be over determined.

In addition we include total variation regularized (TV) reconstructions to compare with the CGLS and non-linear optimization methods. The method employed here is based on the primal-dual algorithm outlined by Chambolle and Pock in \cite{Chambolle2011}, which solves the following optimization problem
\begin{align*}
\arg\min_{\mathbf{m}} \frac{1}{2}\|PR\mathbf{m}-\mathbf{b}\|^2 + \alpha\|L\mathbf{m}\|_{1,2}.
\end{align*}
Here, the mixed $l_1-l_2$ norm $\|\cdot\|_{1,2}$ is defined as 
\begin{align*}
\|L\textbf{m}\|_{1,2} = \sum^{K^2}_{i=1}\|(L\mathbf{m})_i\|,
\end{align*}
where $(L\mathbf{m})_i$ is the forward-difference approximation of the gradient at voxel $i$. The implementation of the primal-dual algorithm also requires four additional parameters that define the primal and dual step-sizes, $\rho,\tau > 0$, and upper and lower bounds for the image domain. The lower bound is known in our case since we cannot have a negative attenuation coefficient, meaning we have three parameters and the regularization factor $\alpha$ to fine-tune for best reconstructions. Despite the required number of parameters and longer runtime due to increased demand for computational power, TV is expected to give more accurate images where objects have distinct boundaries and homogeneous interiors. For this reason TV regularization is often used as a deblurring or denoising method in image processing.

The non-linear reconstructions are obtained using the trust region method \cite{ColemanLi1999} with the reflective transform \cite{Coleman1994} applied at each iteration. The trust region methods minimize a quadratic model function, $Q(\mathrm{m})$ given by
\begin{align*}
\min_{\mathbf{m}} Q(\mathbf{m}) =  f_k + J^T_k\mathbf{m} + \frac{1}{2}\mathbf{m}^TH_k\mathbf{m},\qquad \text{s.t. } \|\mathbf{m}\|\leq \Delta_k,
\end{align*}
where $f_k = \|F(\mathbf{m}_k)-\mathbf{d}\|^2$ is the value of the objective function at iteration $k$, $J_k = D f(\mathbf{m}_k)$ is the Jacobian, $H_k = D^2 f(p_k)$ is the Hessian matrix, and $\Delta_k$ is the radius of the trust region, over which the function $Q(\mathbf{m})$ is minimized. In our implementation, we use the approximation $ J^T_kJ_k$ to $J_k$ (trust region Newton's method). $\Delta_k$ is calculated at each iteration, which is based on the agreement between the quadratic function $Q$ and the objective function $f$ at the previous iteration. See  \cite{ColemanLi1999,Coleman1994,NocedalWrightBook} for implementation details. We chose to implement a trust region method as these methods are more robust than line search methods for solving ill-conditioned systems. 

\section{Description of apparatus}
\begin{figure}\centering
\includegraphics[scale=0.75]{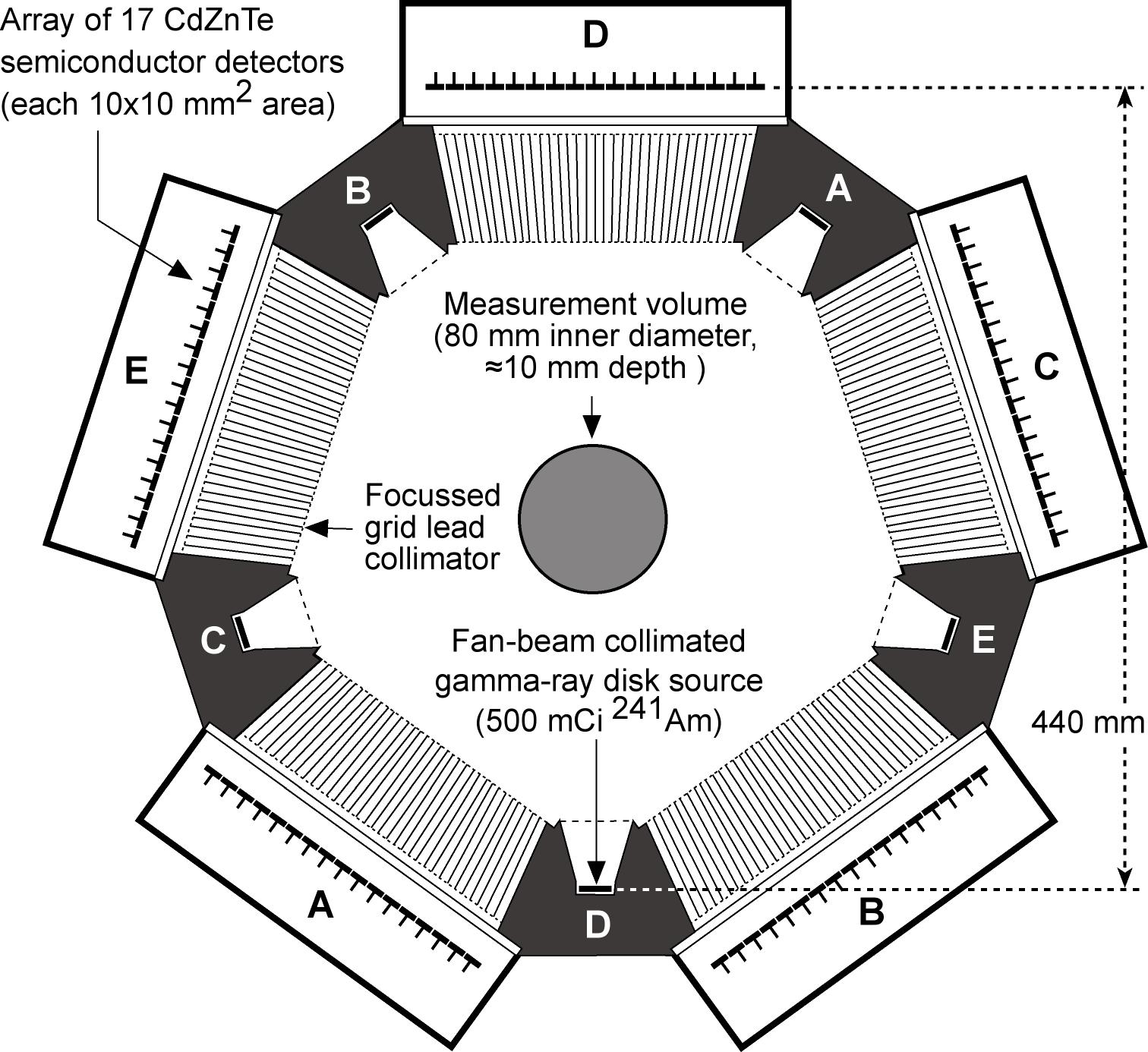}
\caption{Cross section of Bergen $\gamma$-tomography system. The lead collimator plates at the midline of the detector elements are shown but they were removed in the detector array used for our experiments leaving only the plates between each element.}
\label{fig:system}
\end{figure}
The Bergen $\gamma$-ray tomography system described in \cite{Johansen1996} was used for this experimental study, see fig.\ref{fig:system}. The system was designed for fast imaging of oil, gas and water cross sectional distributions within a pipe. The system is made up of a total of 5 $ ^{241}$Am radioisotopes with an activity of 500 mCi and 5 detector modules each containing a total of 17 CdZnTe detectors. The relatively large source (18mm diameter) and detector size (10mm across) separated by 440mm was large enough to expect an observable non-linearity. The $ ^{241}$Am source with a 59.5 keV principal emission peak is a good approximation to monochromatic radiation.

The CdZnTe detector elements are $10 \times 10$ mm$^2$ in cross section and 2mm thick, chosen to effectively stop any low-energy $\gamma$-ray from being transmitted through an element. Lead plates are fitted between each element to reduce the detection of scattered photons: a `scatter grid'. In its normal operation there are similar lead plates bisecting each element but these were removed as we wanted to measure the effect of all the ray paths between source and detector element. These collimation plates would be expected to reduce the non-linearity at the expense of also reducing the counts measured.   Each detector array is positioned opposite a source. The source aperture is designed to collimate the radiation in a fan beam that coincides with the 80mm diameter circular aperture where the phantoms are placed. This aperture is itself surrounded by a tube  of Perspex 5mm thick. The system has five nominally identical source and detector array pairs distributed equiangularly. The counting threshold of each detector is set to 48 keV, effectively screening out the lower emission peaks of the source as well as minimizing the effects of scattered radiation, e.g. those photons that undergo Compton scattering and lose a certain fraction of their initial energy. Before inserting a phantom a background reading was obtained, averaging the counts on the detector array in use until the ratios of the number of counts in each element to the total counts was constant to three decimal places. Each measurement of the phantom used the total number of counts over the same time. In practice we averaged over 30s for each measurement.



\section{Monte Carlo simulation} \label{sec:MC_sim}

It is a well-known fact that scattered radiation 
affects the accuracy of reconstructed images. This is also the case in the Bergen $\gamma$-ray tomography system with fixed source and detector modules as illustrated in fig.\ref{fig:system}. The question that arises then is the following: What is the contribution from scattered radiation to the overall non-linearity as compared to radiation directly transmitted through the phantom and onto the detector module? 
This question cannot be answered easily by experiments as it is not an easy task to separate scattered (i.e. collided) and uncollided components. Monte Carlo (MC) simulations, on the other hand, provide a means to numerically track individual photons in the geometry of the $\gamma$-ray tomography system, and provide the possibility of identifying these as collided and uncollided photons. Thus, in order to study the relative importance of the non-linear effects as compared to scattered radiation, a simple Monte Carlo (MC) model of the Bergen $\gamma$-ray tomography system has been implemented using the general purpose MC code, MCNP6.1 \cite{Goorley2012}.

Here, the first task was to benchmark the MC modeling against experimental data. Since the focus was on determining the overall non-linear effects in the pertinent $\gamma$-ray tomography system, a detailed implementation of entire system geometry was deemed unnecessary. This is because the non-linear effects stem from the partial coverage of an object spanned by the rays between a source and the corresponding detector. A portion of these rays may undergo different amounts of attenuation. Thus, the non-linear effects are not related to the finer details of the measurement geometry with the exception of collimation grid.  
 Therefore, only a single detector module, containing a total of 17 CdZnTe semiconductor detectors, was implemented. In addition, the lead collimator blades placed in front of each detector module were implemented. 
 In order to replicate the experimental conditions, every second collimator, the ones in the middle of each element, was removed from the geometry. Thus, the final geometry implemented in MCNP6.1 contains 17 lead collimator blades. As the main purpose of the MC simulations has been the separation of collided and uncollided fluxes incident upon the detectors, no other details such as the detector and source housing were implemented. In accordance with the geometry of the experimental setup the source-to-detector distance was set to 440 mm. The source was defined as a disk source with a diameter of 18 mm emitting 59.5 keV photons whereas the pipe inner and outer diameters were set to 80 mm and 90mm, respectively. In table \ref{tab:table1}, a list of materials used in the MC simulations is provided.

\begin{table}[h!]
	\centering
	\caption{Listing of materials used in the MC simulations.}
	\smallskip
	\begin{minipage}{12.0cm}
		\begin{tabular}{ p{3.5cm}c c p{1cm}c p{1cm}c }
			\hline
			Component & Elements & Weight fractions & Density [g/cm$^3$] \\ \hline 
			Pipewall (PMMA) & H, C, O  & 0.08, 0.6, 0.32 & 1.19  \\ 
			Phantom(polypropylene) & H, C & 0.14, 0.86 & 0.92  \\ 
			Air & N, O, Ar, C & 0.755, 0.232, 0.012, 0.001 & 0.0012 \\
			Detectors(CdZnTe) & Cd, Zn, Te & 0.4, 0.1, 0.5 & 5.78 \\
			Collimators & Pb & 1.0 & 11.35 \\
			 \hline
			\label{tab:table1}\par
		\end{tabular}
		\vspace{-0.75\skip\footins}
		\renewcommand{\footnoterule}{}
	\end{minipage}
\end{table}

At the energy of interest, the CdZnTe detectors exhibit nearly 100.0$\%$ detection efficiency. Thus, secondary electron transport could be safely neglected. The secondary electron transport was turned off in the entire problem geometry and the simulations were run as a \lq\lq photon-only" problem using detailed photon transport physics. Furthermore, the transport cut-off energy for photons was set to 48 keV in the entire geometry. This is in line with the experiments as the counting threshold of all detectors is set to 48 keV. For all 17 detectors in the geometry, so-called $f4$ tallies giving the flux averaged over a cell were defined. For the purpose of benchmarking the MC model, simulations were run with the pipe filled with air and polypropylene (see table \ref{tab:table1}) representing empty and full pipe measurements, respectively.

Moreover, the above mentioned $f4$ tallies were divided into collided and uncollided contributions using the so-called uncollided secondaries card in conjunction with special tally treatment \cite{Goorley2012}. The purpose was to identify the number of collisions source particles suffer prior to reaching the cells defining the detectors and also, to identify secondary particles such as characteristic X-rays that may reach the detectors without suffering collisions. The uncollided secondaries card thus was used to label these secondaries as collided particles. In this way, it was possible to separate the uncollided source particles from the collided in addition to any secondaries that may reach the detectors. Everything other than uncollided source particles were treated as collided. 

Using the above mentioned settings, the effects of scattered radiation could be studied and compared to the case of \lq\lq no scatter". The results obtained are given in Sec.\ref{sec:res_MC}.

\section{Phantoms and experimental protocol}
\label{sec:phan_exp_protocol}


The biggest observable non-linear effect is expected when the rays between a source-detector pair encounter the largest contrast in the line integral of the linear attenuation. One simple way in which this happens is a plane interface between air and a more attenuating substance with the plane aligned with the source detector pair but only partly in the path. Such a phantom was already available and had been used with the Bergen system. It consist of half of a solid, 80mm diameter cylinder of polypropylene truncated by an axial plane through its mid-line. This was designed to emulate a pipe half full of oil and locating such an interface accurately within the intended use of the system. In each of the five detector arrays the 9$^\text{th}$ of the 17 detector elements diametrically opposite the centre of a source and we thus expect this to show the largest non-linearity. For comparison a full solid cylinder of the same polypropylene was used. The protocol used was to collect a set of data with the half-cylinder phantom aligned with one source and the opposite detector element 9. The phantom was then turned through half a turn and the measurement repeated. Finally the same measurement was repeated with the solid cylinder phantom. To avoid statistical fluctuations due to the arrival time statistics of the photons data was collected over a 30s period in three blocks of 10s. 

Denote the array of normalized logarithmic counts for a each detector element and and a given source  for the half-cylinder in the two positions by $L_1$ and $L_2$ and the full cylinder by $L_{12}$. The normalization is taken with all phantoms removed and compensates for variations in the sensitivity of the detector elements. If the logarithmic counts were linear in the linear attenuation we would satisfy the superposition principle $L_1+L_2=L_{12}$. We thus consider $L_1+L_2-L_{12}$ as a measure of the non-linearity.

\begin{figure}\centering
\subfigure[Numerical phantom used for simulated data.]{\includegraphics[scale=0.63,trim=2cm 1cm 2cm 1cm, clip=true]{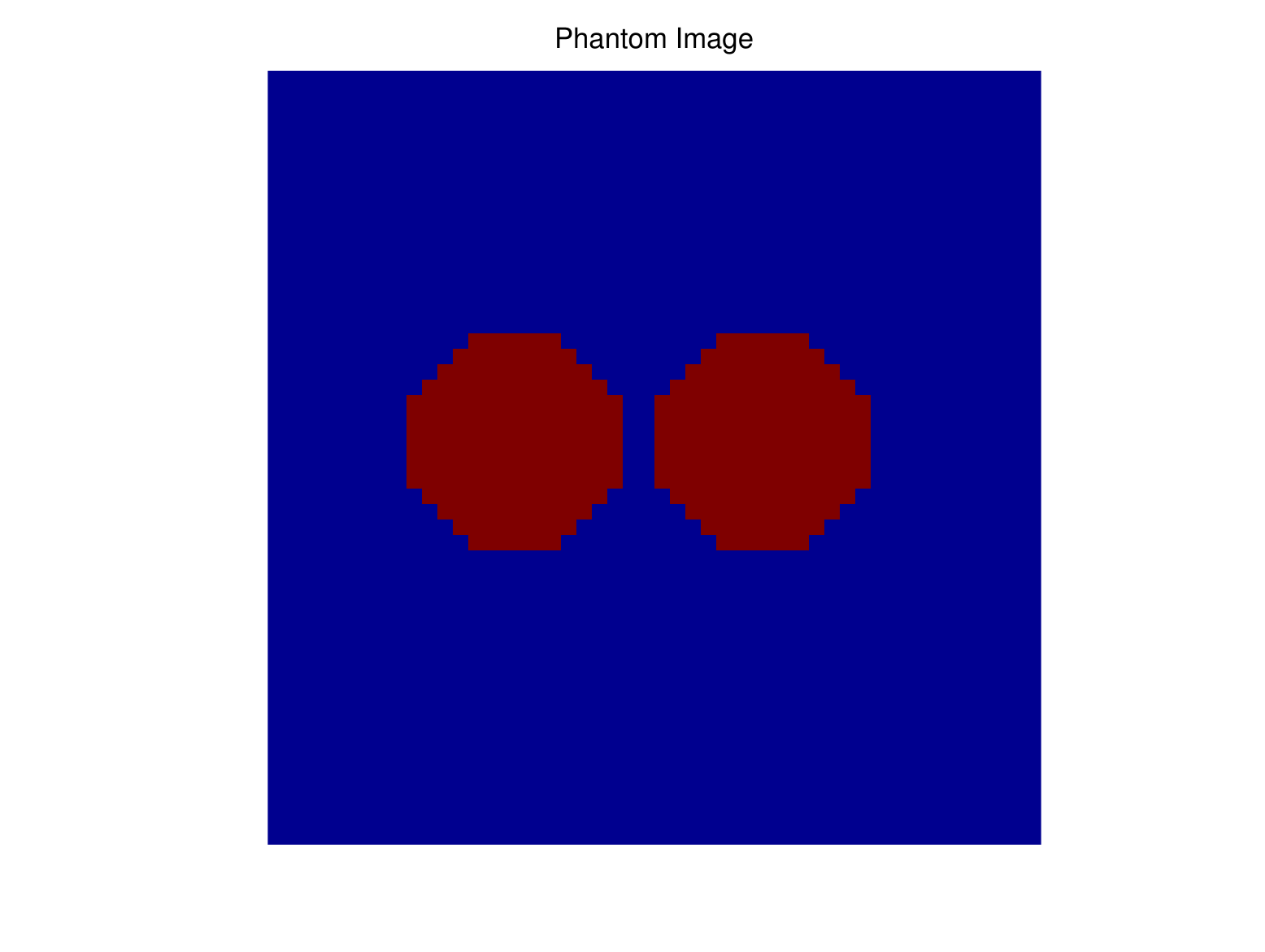}}
\subfigure[Diagram of the real phantom used in $\gamma$-ray experiments.]{\includegraphics[scale=0.515]{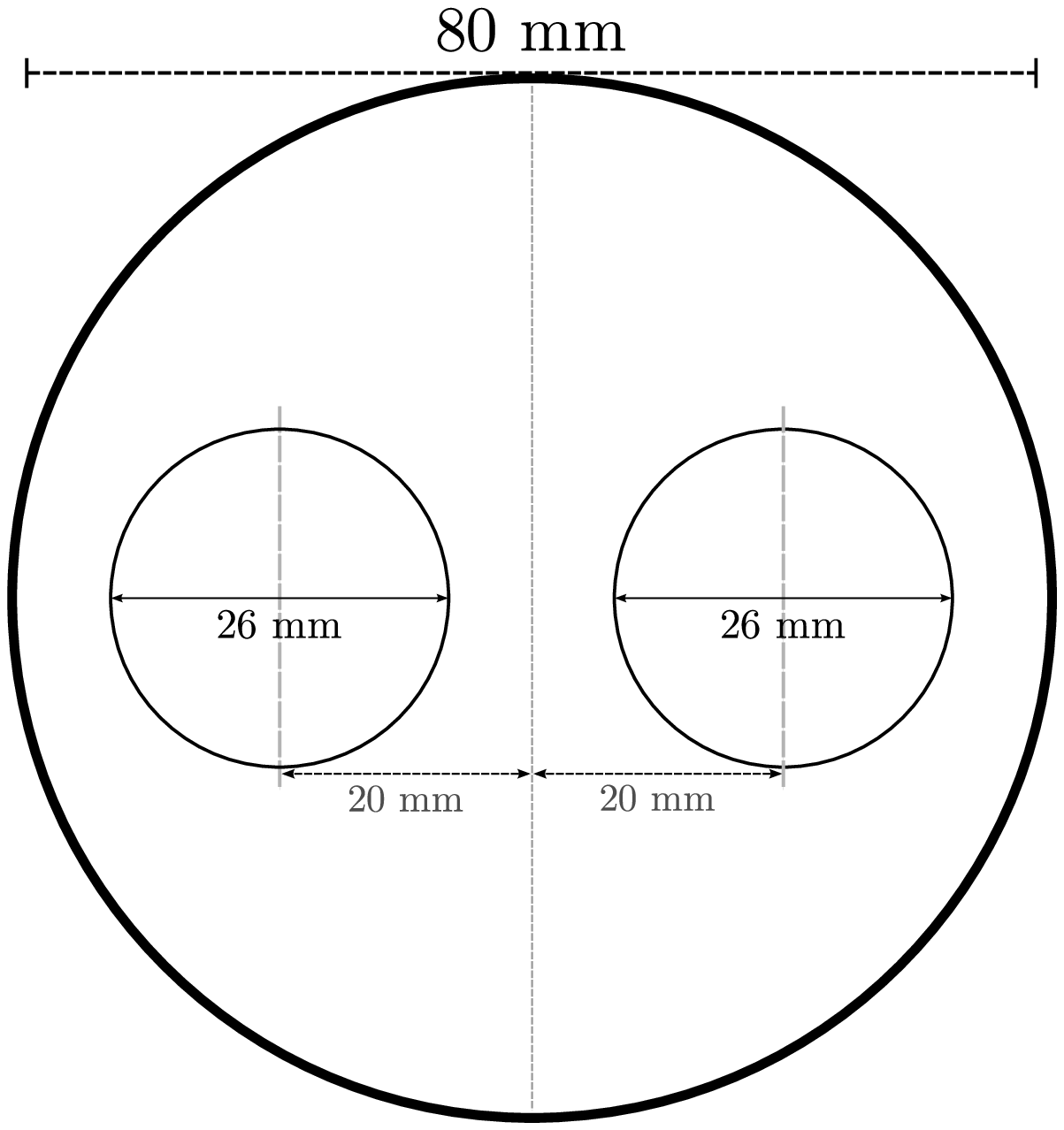}}
\caption{Phantom images used in the simulations (left) and the diagram of the phantom object used for collecting real data (right).}
\label{fig:phantoms}
\end{figure}

The second phantom used consists of an 80mm diameter polypropylene cylinder with cylindrical holes of diameter 26mm bored so that the axis of the cylinder and the hole are parallel and separated by 20mm. See fig.\ref{fig:phantoms}b. Two solid plugs machined to be a tight fit in the hole were made from the same batch of polypropylene. A solid cylinder would have served the same purpose as the two hole cylinder with both plugs fitted, but experiments showed a difference between the linear attenuation factor of our solid cylinder and the polypropylene available to construct the new phantom. The design and dimensions of this phantom were typical of those used to test the Bergen combined $\gamma$-ray and capacitance dual sensor system. Of course we could explore non-linearity using just the plugs suspended in air, but as the apparatus is arranged for a vertical imaging plane it would  be more difficult to ensure the plugs were positioned accurately. Let $L_0$ denote the logarithmic counts for the two hole phantom and $L_1$ and $L_2$ the logarithmic counts with one of the plugs in a hole and $L_{12}$ for both plugs. If the system was linear we would expect $L_1 -L_0$ to be equivalent to the normalized counts for just one plug in air. So superposition would give $L_1 -L_0 + L_2 -L_0= L_{12} - L_0$ so we have $ L_1  + L_2 - L_{12} - L_0$ as a measure of the non-linearity.

We also collected data suitable for comparing linear and non-linear reconstruction from multiple projection. The Bergen system is intended to be used for fast imaging using data from only five projections, collected simultaneously from five sources. To remove the limited data effect of only five projections and to eliminate variation caused by differences in the sources and detector arrays we collected imaging data by rotating the half cylinder and two hole phantom in angular increments to collect a fully sampled fan-beam projection data set. Again, the measurement time for each projection was 30s and thus, the statistical fluctuations in the data were negligible.

Finally our initial numerical experiments used a numerical phantom, see fig.\ref{fig:phantoms}a. This consisted of a square pixel grid with a background of $\mu=1$ with circular contrasting objects with $\mu=2$. The dimensions of the square were taken so that the diagonal was contained in the 80mm region of interest. To avoid `inverse crimes' the  simulated data was generated using a grid with $2K \by 2K$  pixels Gaussian pseudo random noise was added to the data.

\section{Results}

\subsection{Results of MC simulations} \label{sec:res_MC}
In this section we present the results of MC simulations. 
Firstly, the MC implementation of the measurement geometry was benchmarked with experimental data. All simulations were run for a total of $10^8$ primary photon histories, ensuring a relative statistical error of less than 2\% for all tallies in the problem, i.e. collided and uncollided fluxes incident on the detectors for empty pipe, full pipe and phantom measurements. Each simulation was run on a six-core Intel Xeon 2.4 GHz CPU and took about 25 minutes to complete.

\begin{figure}[ht!]
	\begin{center}
		\includegraphics[scale=0.5]{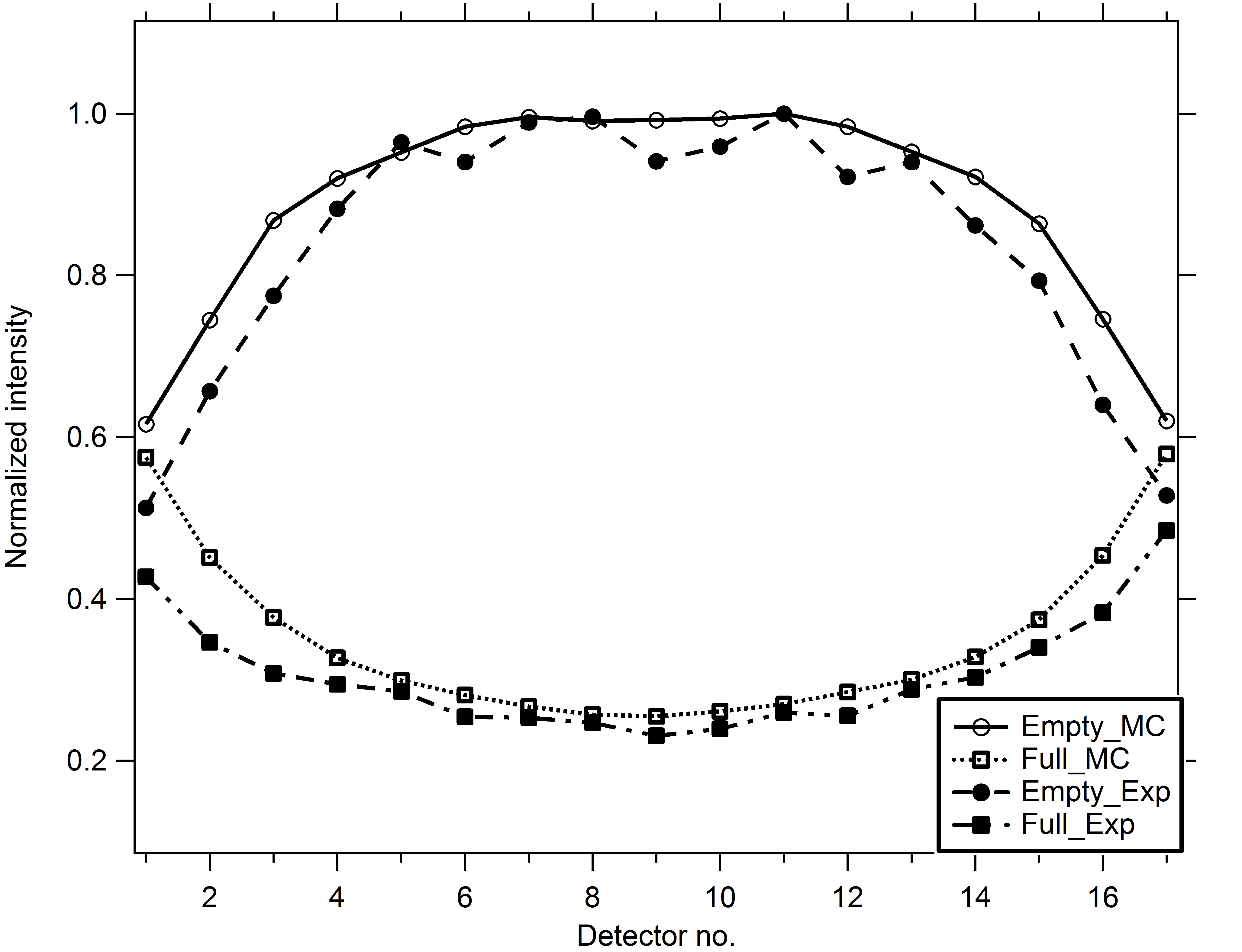}
	\end{center}
	\caption{\label{fig:MC_benchmark}The normalized intensities recorded by each detector for empty and full pipes. The open legends show the results of MC simulations whereas the full legends show the experimental results. The normalization of the intensities is performed by dividing the intensities in each detector by the largest intensity obtained in the empty pipe measurements.}
\end{figure}

As can be seen in fig.\ref{fig:MC_benchmark}, the results of MC simulations showing the total flux of photons incident on the detectors agree reasonably well with the experimental data, especially for the central detectors in both full and empty pipe measurements. The MC simulations and experiments does, however, show a divergent behaviour for the side detectors. For the side detectors, as fig.\ref{fig:MC_benchmark} reveals, the MC simulations predict higher intensities. This may be explained by considering the following factors; finer details of the measurement geometry are ignored in the MC modeling, only nominal values of the pipe dimensions are considered and positioning of the lead collimators. MC simulations do confirm this, as e.g., using slightly larger pipe diameter improves the agreement between MC simulations and experiments. As given in the following, altering these dimensions does not change the results significantly with respect to the non-linear partial volume effects vs. scattered radiation. Thus, for the purposes of this work, it was concluded that the MC simulations, given the results shown in fig.\ref{fig:MC_benchmark}, gave a reasonably well approximation of the experimental apparatus.    

The non-linear effect was studied using a half-circular phantom simulating a pipe half oil filled and half empty (i.e., air). As given in Sec.\ref{sec:resexp}, the experiments show a significant non-linear effect at the interface between phantom and air whereas the largest non-linearity is seen for the center-most detector, i.e. detector 9, which spans this interface. It would therefore be interesting to see whether this effect would be equally pronounced when there is no scattered radiation. The MC predicted photon flux incident on the detectors were separated into collided and uncollided components. As mentioned earlier, collided component includes contributions from both scattered primary photons as well as any secondary photons from atomic relaxations. The non-linearity was calculated for the uncollided flux as outlined in Sec.\ref{sec:phan_exp_protocol}.

  \begin{figure}[ht!]
  	\begin{center}
  		\includegraphics[scale=0.5]{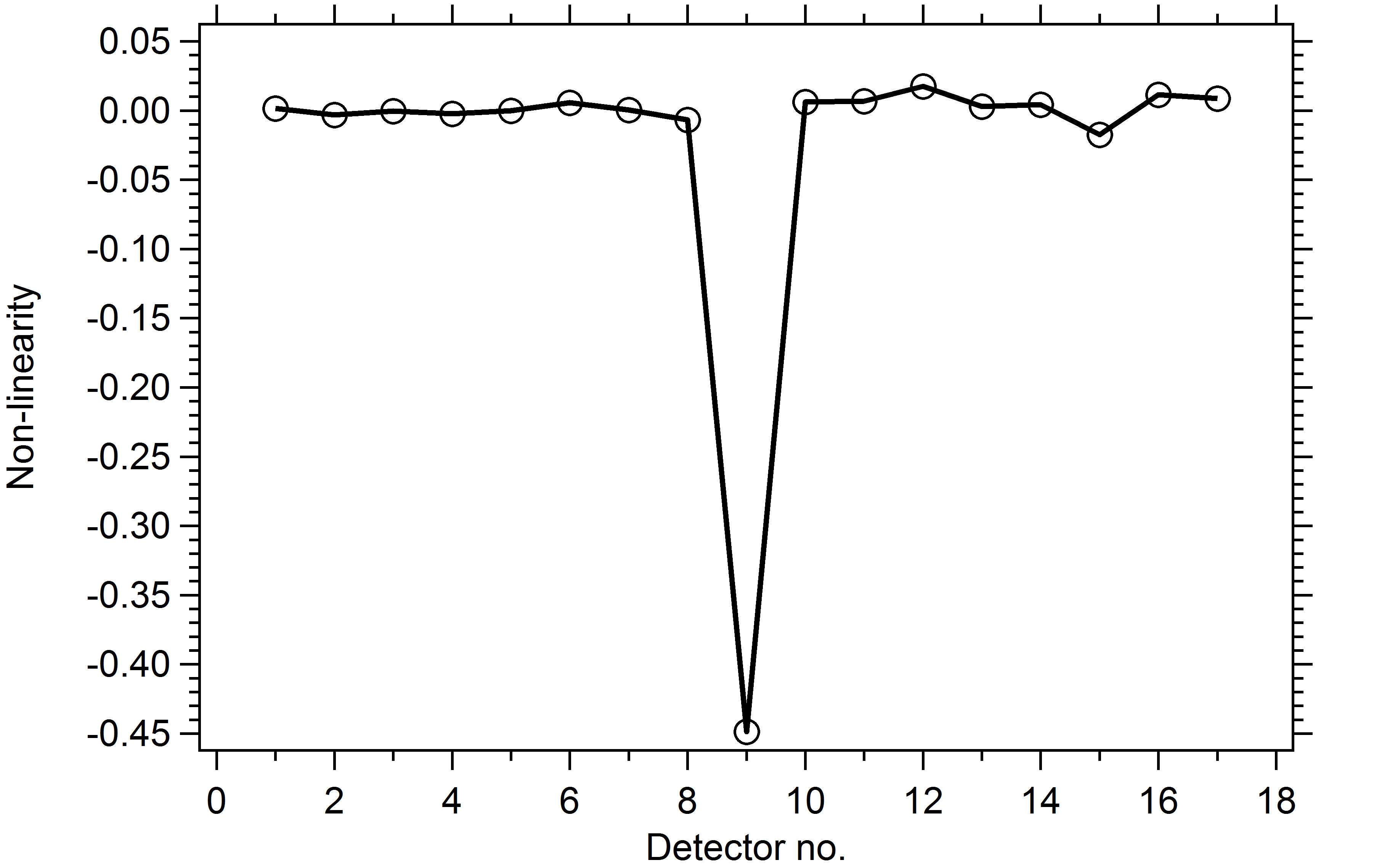}
  	\end{center}
  	\caption{\label{fig:nonlinDMC}The MC generated non-linearity plotted against the detector number in a given detector module obtained for the half-circular phantom using only the uncollided photon flux. The results show that the non-linearity will be substantial (for the center-most detector, i.e. detector 9) even in case of no scattered photons.}
  \end{figure}

As expected, and as shown in fig.\ref{fig:nonlinDMC}, non-linearity turns out to be largest for the centre-most detector, i.e. detector 9, in this detector module. More importantly, the benchmarked MC results show that the non-linearity will be substantial, even when scattered radiation is entirely eliminated.

The above observation partially confirms the statement that the non-linear partial volume effects are present even when the scattered radiation is eliminated entirely and that the non-linearity in the transmitted radiation is the dominant cause of errors in images reconstructed assuming a linear forward model. A direct comparison based on the use of logarithms of collided and uncollided intensities is, however, difficult as the normalization of intensities is not straightforward. In addition, the overall intensity of scattered radiation is, in the given geometry, an order of magnitude less than the uncollided intensity. 

As only one source-detector module pair was implemented, full and empty pipe simulations were repeated five times using different random number seeds to obtain all five projections. The same procedure was utilized for the simulation of the phantom whereas the phantom was rotated at intervals of 72$^\circ$. The tallied photon fluxes were separated as total (i.e., sum of collided and uncollided fluxes) and uncollided. 

\begin{figure}[ht!]
	\begin{center}
		\includegraphics[width=\textwidth]{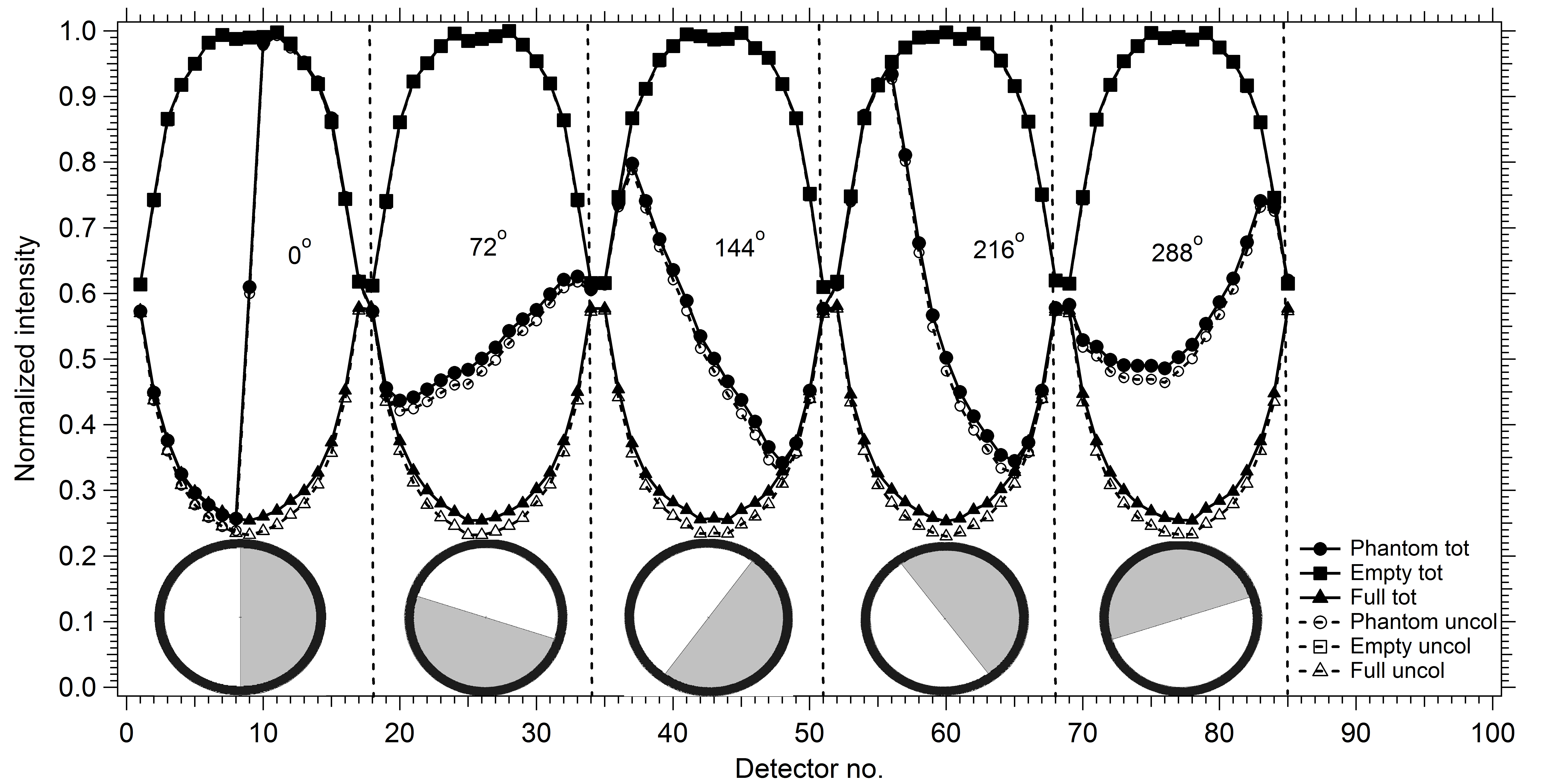}
	\end{center}
	\caption{The MC generated, normalized intensities shown for both total (i.e., sum of collided and uncollided) and uncollided fluxes. Open legends and dashed lines show the intensities for the uncollided fluxes whereas the full legends and solid lines show those of total fluxes.}
    \label{fig:simintensities}
\end{figure}

In fig.\ref{fig:simintensities}, the MC simulated, normalized intensities obtained for the two cases, i.e. total and uncollided fluxes, are shown along with the rotation of the half-circular phantom. It should be kept in mind that, to accelerate the simulations, a photon transport cut-off energy of 48 keV was applied in all of the simulations. Thus the results reflect the intensities obtained for photons arriving at the detectors with energy greater than 48 keV. As mentioned earlier, this is in line with the experiments as a counting threshold of 48 keV is applied to each detector in the $\gamma$-ray tomography system.    

These results support the statement that scattered radiation only has a marginal effect as compared to the non-linear partial volume effects. It should, however, be emphasized that effects of scattered radiation will be dependent on the measurement geometry, more specifically on the source-detector separation as well as the collimation grid used to reduce the effects of scattered radiation. The geometry of the Bergen $\gamma$-ray tomography system is optimized to minimize the flux of scattered radiation incident on the detectors. The source-detector module separation is large compared to the pipe dimensions, the detectors are heavily collimated using lead blade collimators and a counting threshold of 48 keV is applied in order to further eliminate the influence of scattered radiation. These are factors that contribute to the fact that scattered radiation intensity in the given geometry is greatly suppressed. Thus, a new set of MC simulations were run where all of the lead collimators were removed and the source-detector module separation was reduced to half the initial value of 440 mm. This increases the flux of scattered radiation incident on the detectors by about a factor of 5. Here, the corresponding intensity profiles are not be shown explicitly as these are essentially similar to the ones obtained for the original system shown in fig.\ref{fig:simintensities} .



\subsection{Results of experiments} \label{sec:resexp}

\begin{figure}[ht!]
  	\begin{center}
  		\includegraphics[scale=0.5]{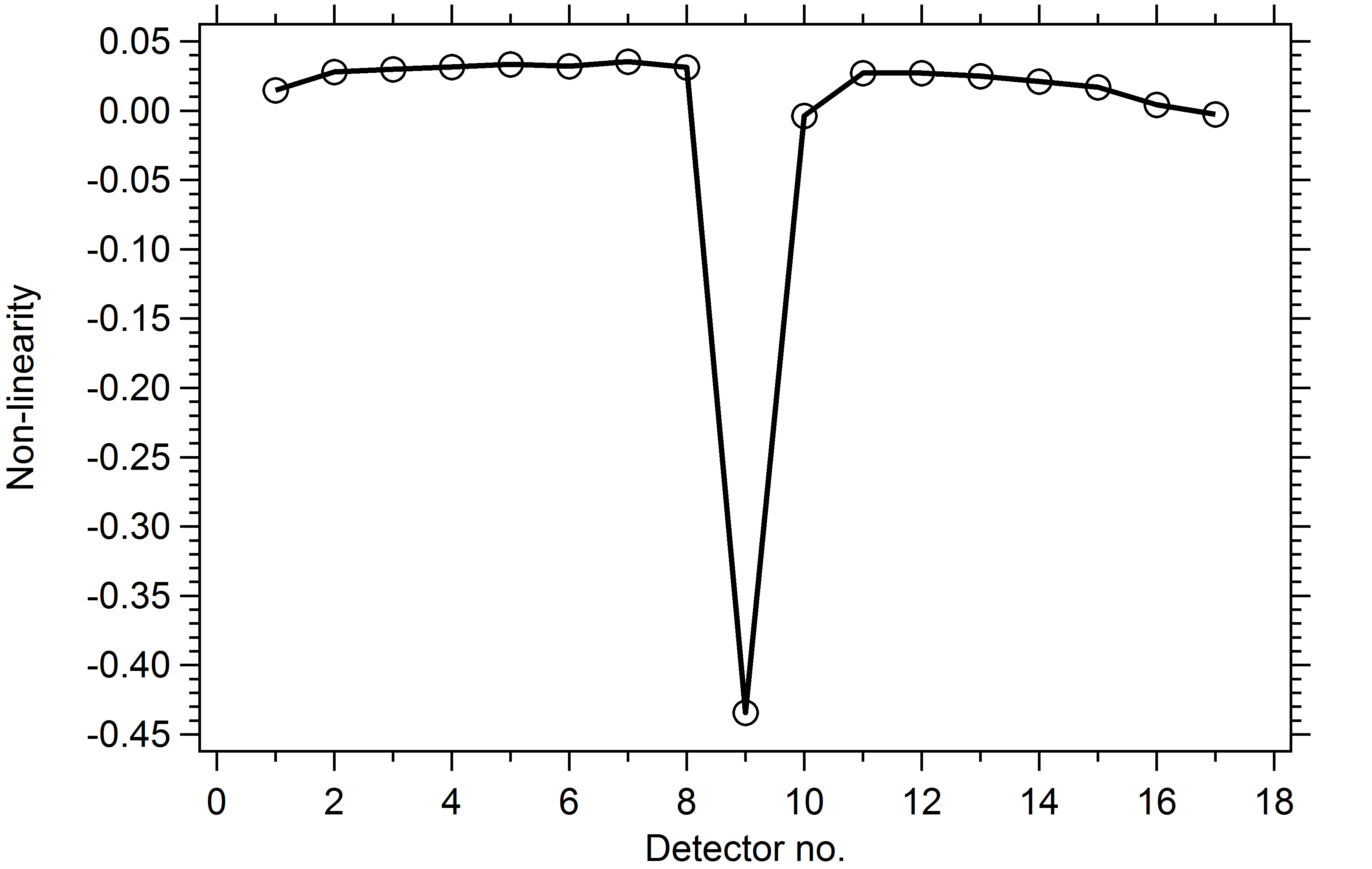}
  	\end{center}
  	\caption{\label{fig:nonlinDExp}The experimental non-linearity plotted against the detector number in a given detector module obtained for the half-circular phantom. The results show that the non-linearity will be highest for the centre-most detector, i.e. detector 9, that spans the interface between air and plastic.}
  \end{figure}

Our initial experiment on the half circular phantom was to determine if there was a significant non-linear effect. The non-linearity was calculated as described in Sec.\ref{sec:phan_exp_protocol} and the results are given in fig.\ref{fig:nonlinDExp}. 

The figure show a graph of non-linearity in the logarithmic data from this experiment. As expected the non-linearity is highest for the source detector pair that spans the interface between air and plastic. As the phantom simulates a half full pipe of oil it was interesting that the non-linearity is significant for this important case.

\begin{figure}[ht!]\centering
\subfigure[CGLS with Tikhonov regularization.]{\includegraphics[scale=0.49,trim=3cm 1cm 3cm 1cm, clip=true]{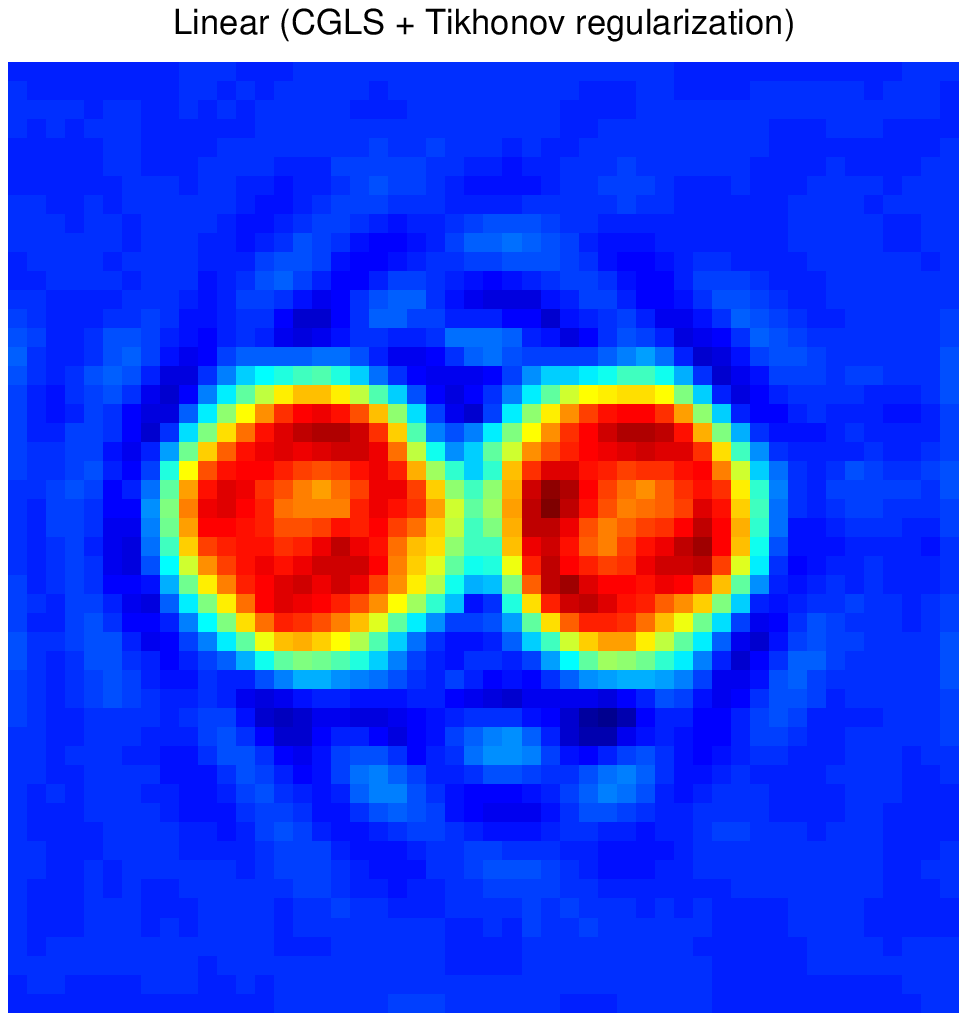}}
\subfigure[Primal-dual with TV regularization.]{\includegraphics[scale=0.49,trim=3cm 1cm 3cm 1cm, clip=true]{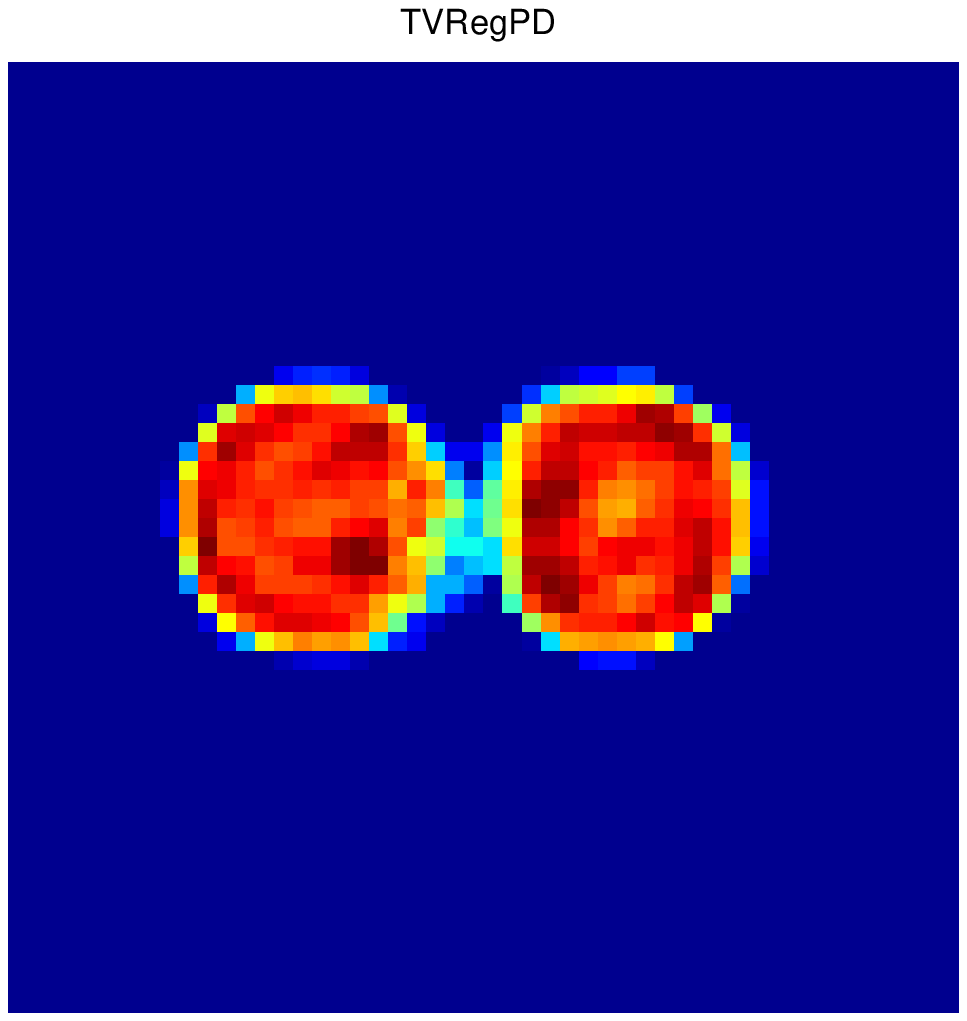}}
\subfigure[Trust region reflective method.]{\includegraphics[scale=0.49,trim=3cm 1cm 3cm 1cm, clip=true]{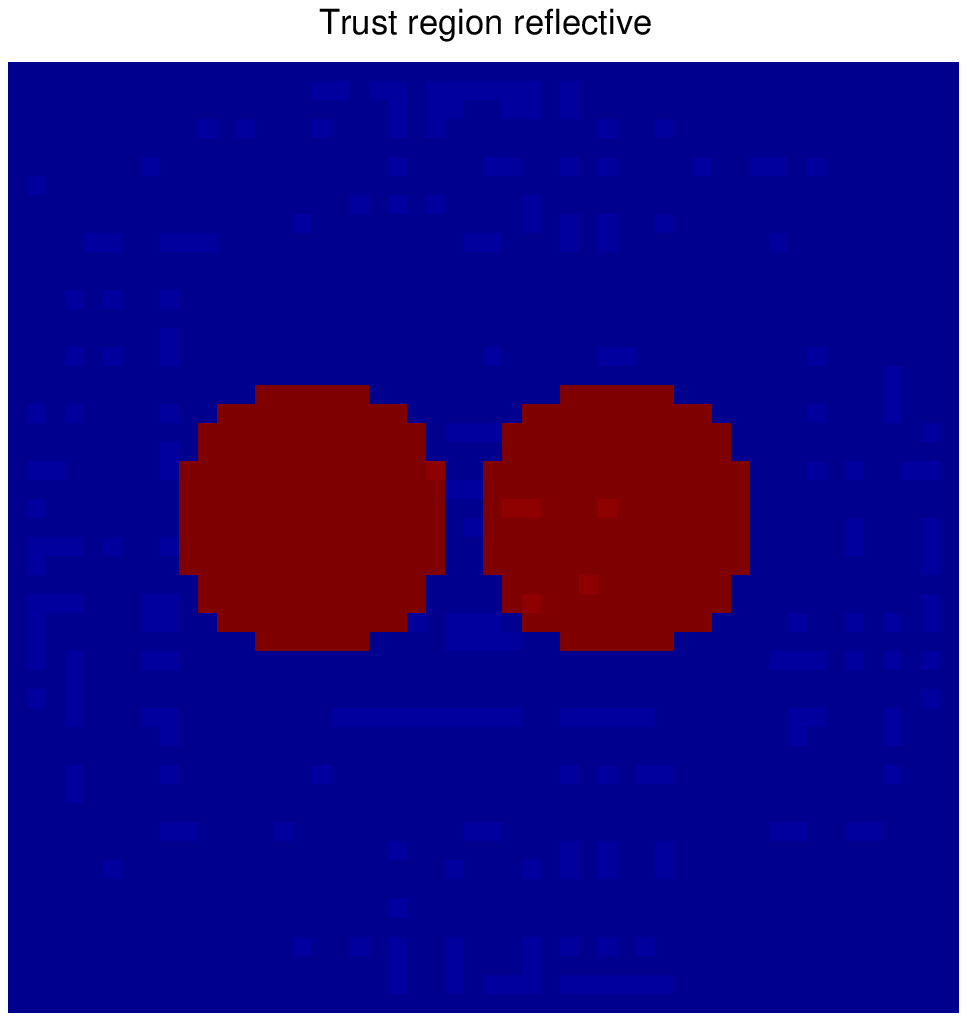}}
\caption{Simulation results for linear and non-linear reconstructions. Simulated data is generated with 17 rays per projection, 180 projections. Images are reconstructed over a $50\by 50$ square grid.}
\label{fig:simple_phantom_180proj}
\end{figure}

We then tested the algorithms described in Sec.\ref{sec:recon_algs} on the data from the numerical phantom with $2^\circ$ rotations, giving 180 projections, and real $\gamma$-ray data with the Bergen $\gamma$-tomography system, also with $2^\circ$ rotations. Fig.\ref{fig:simple_phantom_180proj} shows the reconstructions obtained via CGLS with Tikhonov regularization, primal-dual with TV regularization and the trust region reflective methods. For the CGLS runs, we chose the tolerance value $\epsilon = 10^{-4}$, and maximum number of iterations to perform to be $i_{\text{max}} = 10^4$ for both numerically simulated and real $\gamma$-ray data reconstructions. The same tolerance value was also used in the trust region reflective method. In addition, TV parameters were fine-tuned separately for simulated and real data experiments: $\rho,\tau = 1.5$, $\alpha= 10^{-5}$, upper bound chosen as 1 for simulated; $\rho,\tau = 1.9$, $\alpha= 10^{-3}$, upper bound $5$ for real $\gamma$-ray data reconstructions. The simulated geometry for these runs was set up to imitate the Bergen $\gamma$-tomography system, namely with 17 rays projected onto a flat array of detectors from a single source. The data was generated using the non-linear forward model applied to the phantom image in fig.\ref{fig:phantoms}a, and 10\% white Gaussian noise was added to the simulated $\gamma$-ray data. Of course this level of noise would represent a much shorter averaging time, in practice closer to how the apparatus is typically used for in-situ measurements. Except that of course in its standard use only five projections would be taken simultaneously.

\begin{figure}[ht!]\centering
\subfigure[CGLS with Tikhonov regularization.]{\includegraphics[scale=0.49,trim=3cm 1cm 3cm 1cm, clip=true]{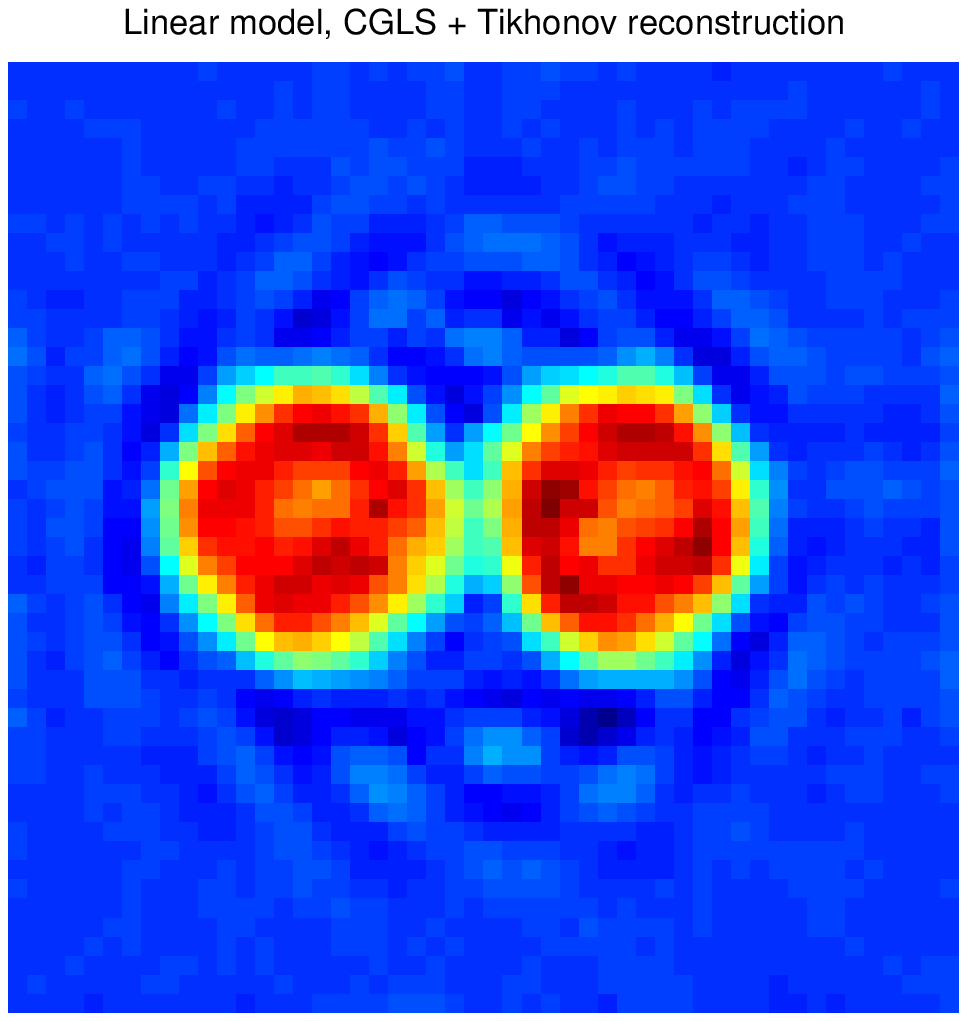}}
\subfigure[Primal-dual with TV regularization.]{\includegraphics[scale=0.49,trim=3cm 1cm 3cm 1cm, clip=true]{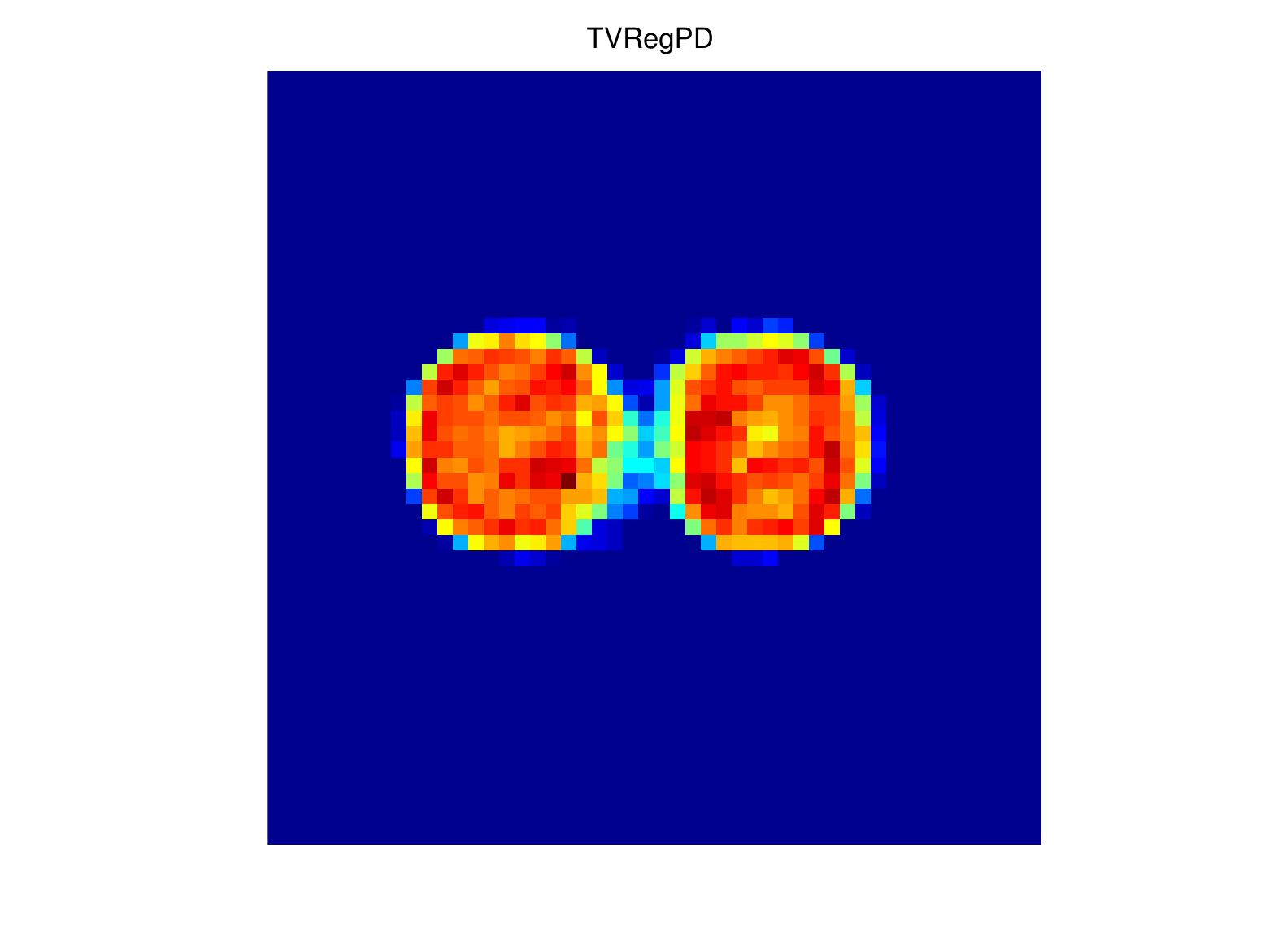}}
\subfigure[Trust region reflective method.]{\includegraphics[scale=0.49,trim=3cm 1cm 3cm 1cm, clip=true]{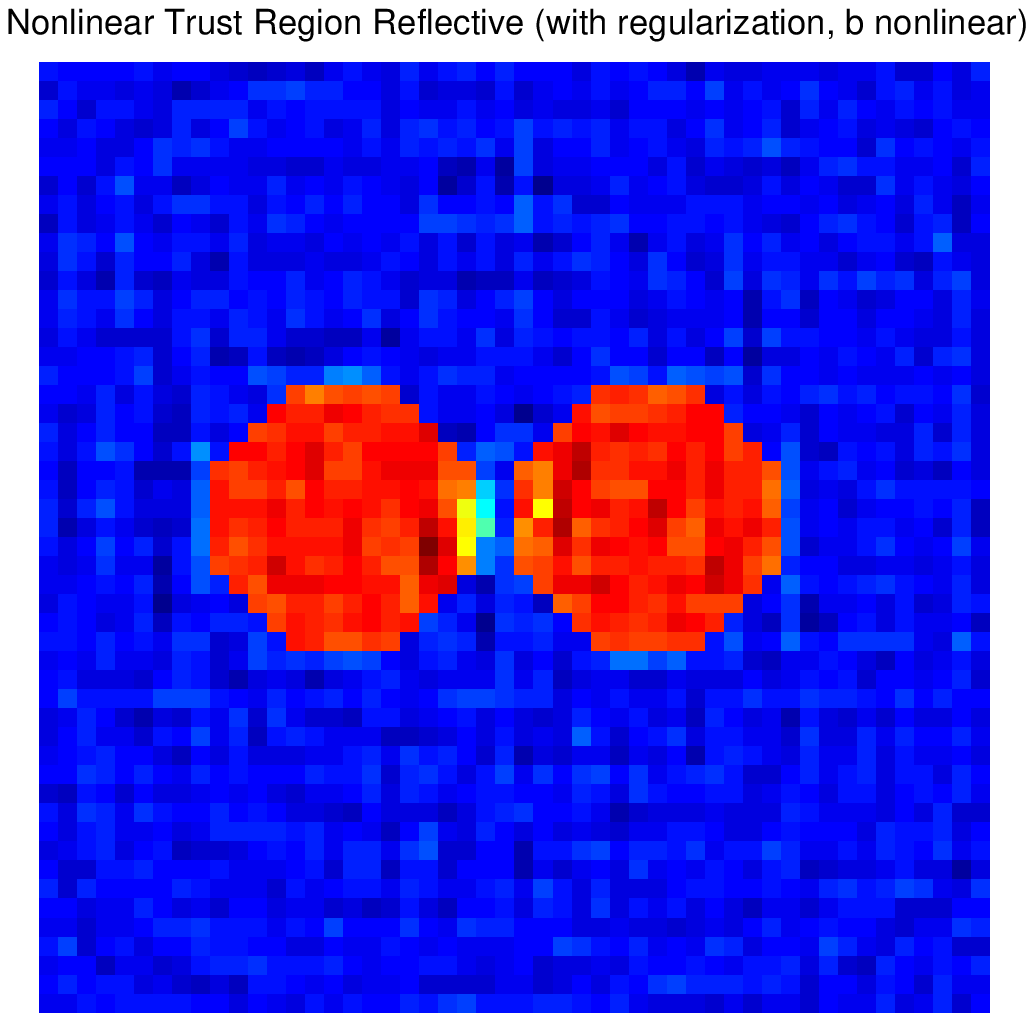}}
\caption{Simulations are repeated with fewer angles (60 projections).}
\label{fig:simple_phantom_60proj}
\end{figure}

The numerical experiment was repeated with simulated data collected over 60 equiangular projections. The results are shown in fig.\ref{fig:simple_phantom_60proj}. The same reconstruction parameters were used as the previous case with 180 projection angles. 

\begin{figure}[ht!]\centering
\subfigure[CGLS with Tikhonov regularization.]{\includegraphics[scale=0.49,trim=3cm 1cm 3cm 1cm, clip=true]{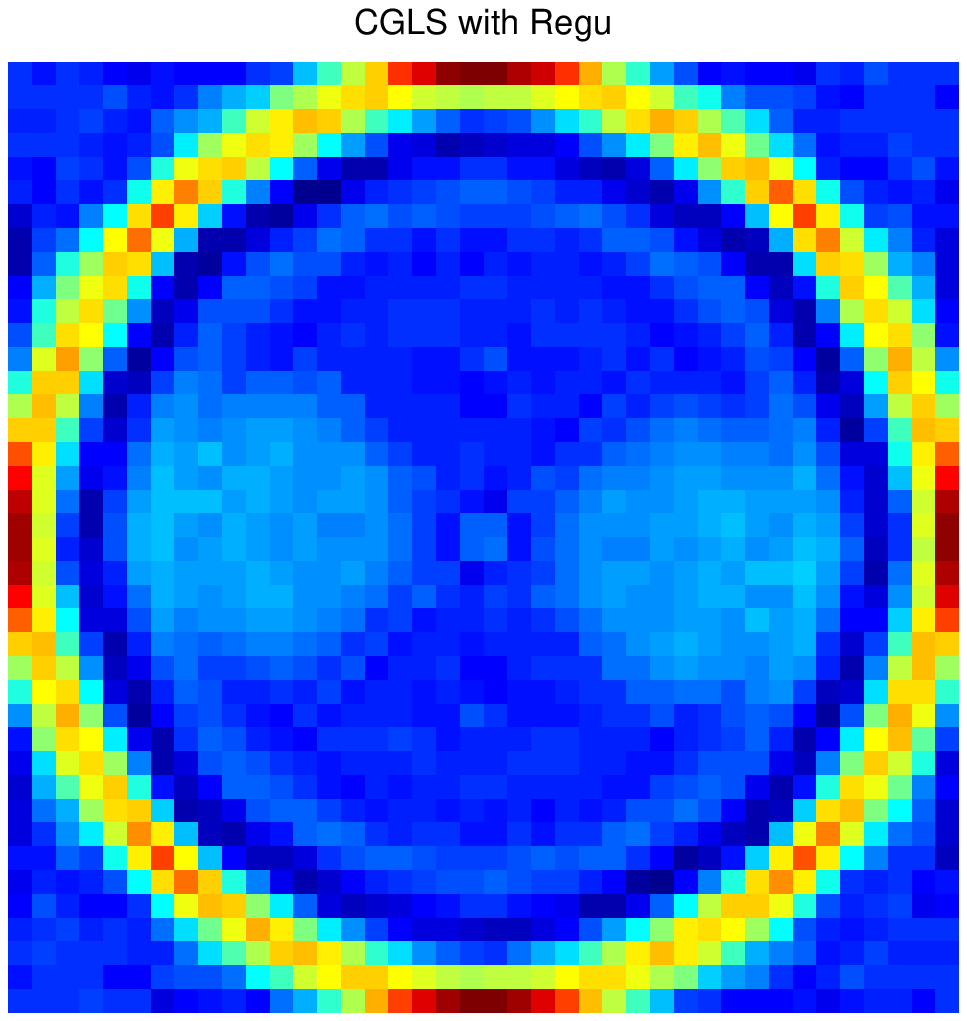}}
\subfigure[Primal-dual with TV regularization.]{\includegraphics[scale=0.49,trim=3cm 1cm 3cm 1cm, clip=true]{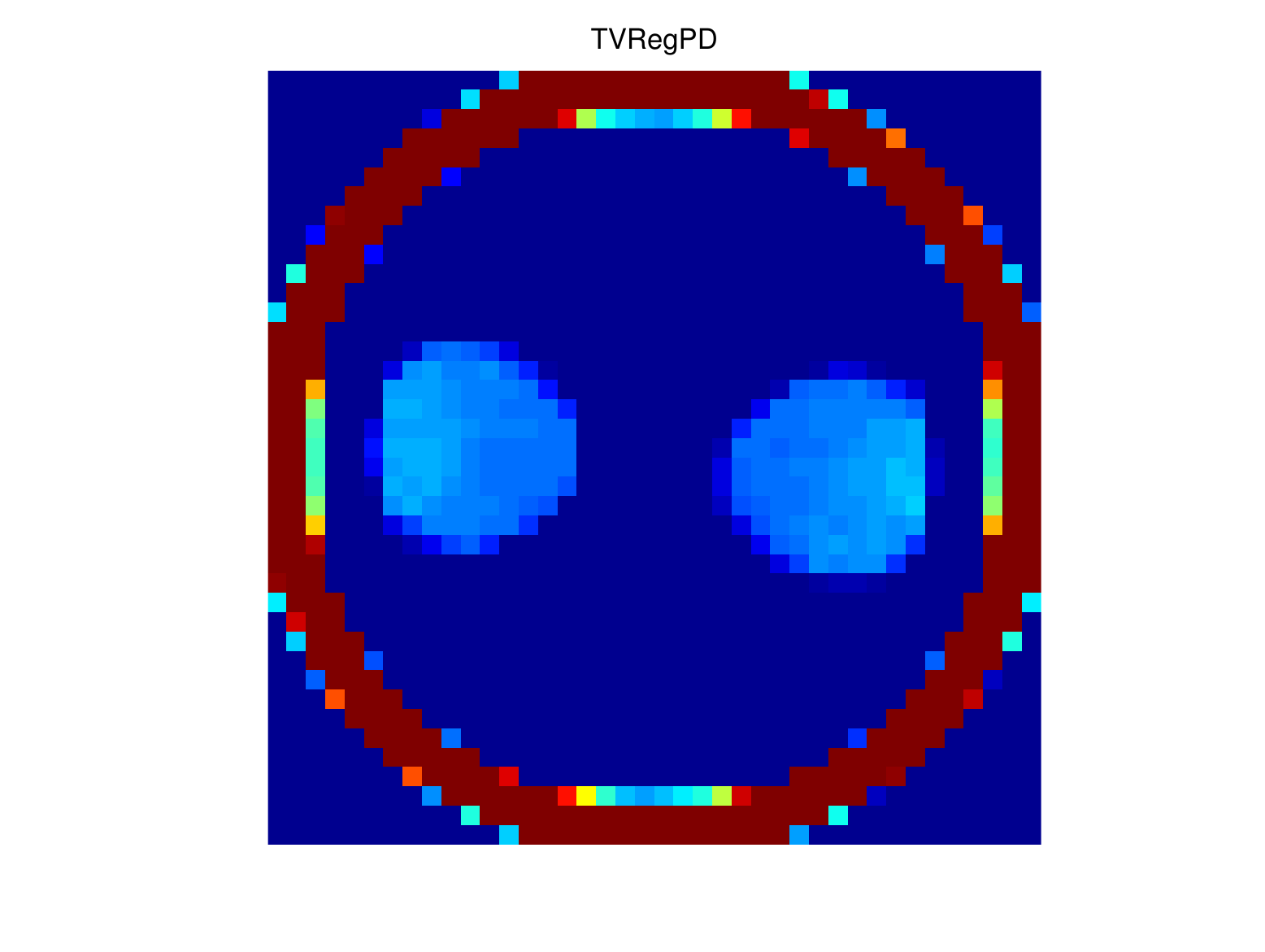}}
\subfigure[Trust region reflective method.]{\includegraphics[scale=0.49,trim=3cm 1cm 3cm 1cm, clip=true]{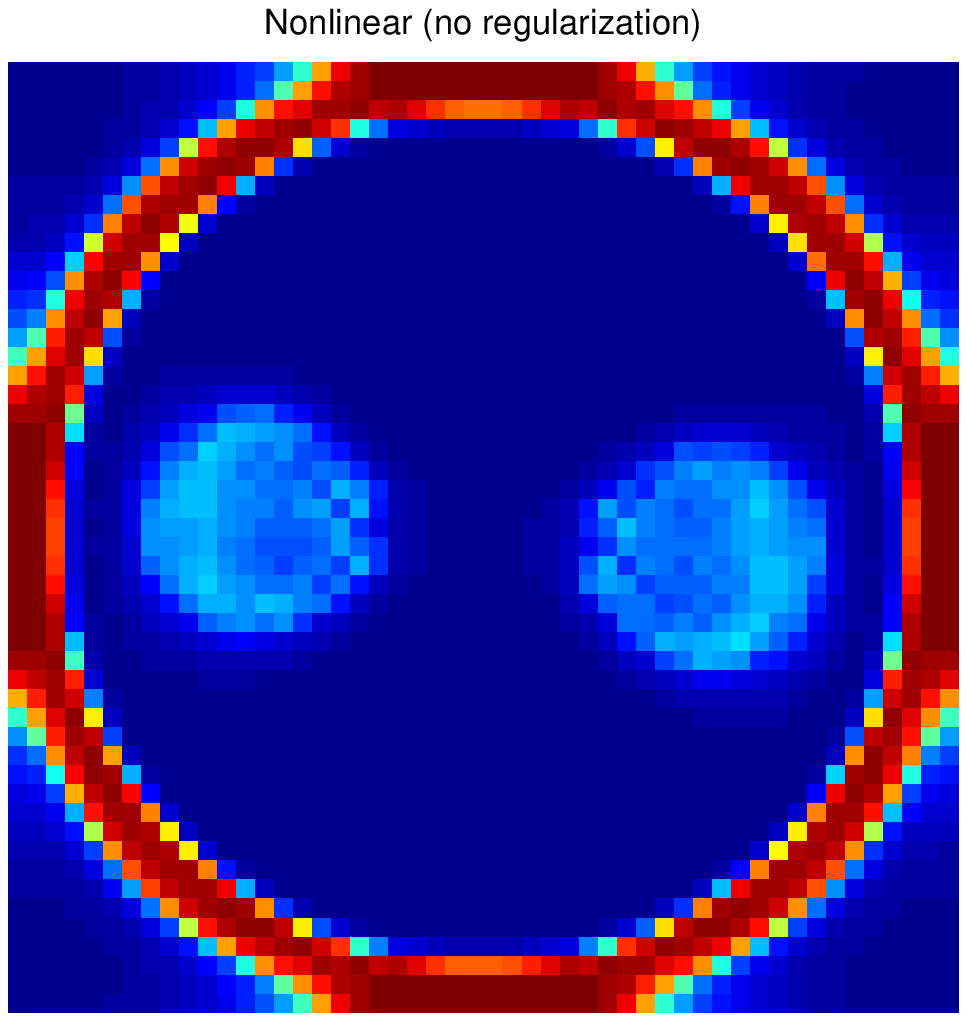}}
\caption{Real data reconstructions using the linear and non-linear models. Provided data contains 17 rays per projection, 180 projections. Images are reconstructed over a $50\by 50$ square grid.}
\label{fig:real_data}
\end{figure}

Finally we present reconstructions from the experimental $\gamma$-ray data collected with the Bergen $\gamma$-tomography system using the linear and non-linear models from the two hole phantom, illustrated in fig.\ref{fig:phantoms}b. The results are shown in fig.\ref{fig:real_data}. 
In all reconstructions the perspex ring is clearly shown and the holes are visible. The CGLS method with Tikhonov regularization produces the poorest image (in quality) while primal-dual with TV and the trust region reflective reconstructions are very close in accuracy. However it should be noted that the over-smoothing with TV regularization is clear within the holes and the perspex ring, and further, over-estimating the thickness of the ring in comparison to CGLS with Tikhonov and trust region reflective methods. This is also demonstrated in the simulated results where the two hexagonal shapes are over-smoothed into more circular discs.  

\section{Discussion and Conclusions}




Computed tomography problem is generally taken as a linear problem in the sense that the logarithm of the collected data on each detector, from each source, is a linear function of the attenuation coefficient. However, as we demonstrated by considering the  Beer-Lambert law, the sum of exponential does not imply the exponential of the sum. This effect, (rarely) noted in the literature as the \lq\lq exponential edge-gradient\rq\rq or \lq\lq non-linear partial volume\rq\rq effect, is magnified for systems with large sources and detectors producing a wider range of ray paths.

In this paper, we demonstrate that this non-linearity is clearly present in a monochromatic tomography system. For what we believe to be  the first time in literature, we perform a non-linear reconstruction taking account of this effect using regularized non-linear optimization methods.

For the reconstructed results presented in this paper, we used the trust region reflective method to solve the non-linear problem by minimization of an objective function. The results were compared with the reconstructions obtained with a popular reconstruction method, CGLS. The stark differences in the CGLS and the trust region reflective reconstructions highlight the dominance of the non-linearity effect around the edges of the objects in the phantoms. This effect is also highlighted in simulated primal-dual TV regularized reconstructions with a linear forward where two hexagonal regions are close to one another. 


These initial reconstruction results show that for sufficiently large sources and detectors taking account of the non-linearity can produce dramatic improvements. However there is still much work to be done. We have solved a non-linear problem where the linearization about the background case is well explored and clearly invertible under the right conditions on the pixel grid. Detailed analysis is required to show that our regularized non-linear optimization problem does not suffer from local minima.  Also we used a standard optimization algorithm with the non-linear forward model and a more customized algorithm for this problem is likely to be more efficient. With the linear forward problem we imposed a smoothness condition on the image; we implemented a more sophisticated algorithm for objects with distinct boundaries and relatively homogeneous interiors, for which total variation regularization gave better images, as did the imposition of strict upper and lower bounds using constrained optimization methods.

In the specific case of the Bergen $\gamma$-ray tomography system, the lead collimators were expected to reduce counts registered due to scattering so we are confident that the non-linearity we observed is largely due to the mechanism we describe. The results of MC simulations of the system confirm this statement. The MC simulations predict that the intensity of the scattered radiation is about an order of magnitude less than the transmitted radiation intensities for the given system. This is not surprising as the system has been designed and optimized to minimize the effects of scattered radiation. The reconstructed images of the MC simulated half circular phantom show also that the error in the calculated GVF based on the reconstructed pixel densities has the same order of magnitude even for scatter-free data. Most importantly, for the half circular phantom, the non-linearity in the centre-most detector is substantial even in case of full scatter rejection, indicating the predominance of non-linear partial volume effects. Through MC simulations, it was also demonstrated that, by removing the collimation grid entirely and reducing source-detector separation, scattered radiation has a more pronounced effect on the reconstruction accuracy. However, the non-linear reconstruction on experimental data as well as absolute deviations from the true GVF calculated for the MC generated total and scatter-free data reveal that it would be necessary to take into account also the non-linearity for a substantial improvement in the reconstruction accuracy. For polychromatic tomography systems, MC simulations would also be able to predict how much of the non-linearity in a polychromatic system is due to the size of the source and detectors as opposed to the variation of attenuation with energy (beam hardening). In X-ray systems the source would be expected to be much less uniform. In \cite{Kueh2016} a Gaussian was used to model the source intensity of the focal spot in a micro-focus laboratory CT system. Such variation of $I_0$ is easily incorporated in our algorithm if it is known.

\section*{Acknowledgements}
WRBL is partly funded by The Royal Society Wolfson Research Merit award, WRBL and SBC would like to thank EPSRC for support from grants  EP/M022498/1, EP/I01912X/1 and EP/M010619/1 and the EU for COST Action MP1207 EXTREMA.   
The design and development of the University of Bergen $\gamma$-ray tomography was funded by the Research Council of Norway, Norsk Hydro ASA (now merged into Statoil ASA) and the University of Bergen. Recent hardware upgrade development work has been funded by GCE Subsea (grant$\#$ 301602). IM would like to acknowledge a former student, IVM Moreira for access to reconstruction software used for reconstructing the MC simulated data.

\bibliographystyle{plain}
\bibliography{nonlinearct}

\appendix
\section{Point sensitivity}\label{AppendixA}

\begin{figure}[ht!]\centering
\includegraphics[width=0.8\textwidth]{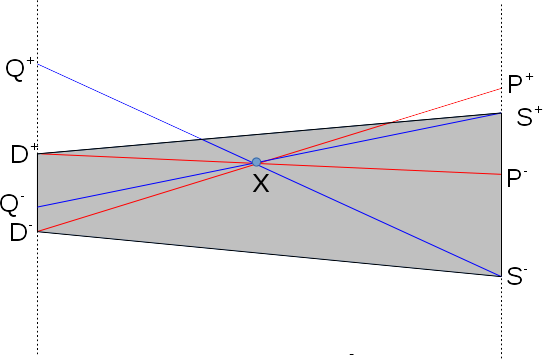}
\caption{Simple two-dimensional case of point sensitivity.}
\label{fig:beamgeom}
\end{figure}

 Consider a two dimensional example where (see figure \ref{fig:beamgeom}) The source is a line segment $S^-S^+$ and the detector is a parallel line segment $D^-D+$ placed opposite the source at a distance $L$. Let $X$ be the point unde linear forward problem we chose the simplest regularization thaer consideration. Consider the lines $XS^-$ and $XS^+$ from the ends of the source meeting the detector line $D^-D^+$ at $Q^-$ and $Q^+$ which may or may not be within the detector. Similarly extend lines $D^-X$ and $D^+X$ from either end of the detector to meet the source line $S^-S^+$ at $P^-$ and $P^+$. The only rays passing through $X$ that result in a measurement are those between points in the intersection of line segments $D^-D^+ \cap Q^-Q^+$ and $S^-S^+ \cap P^-P^+$ so the point sensitivity is proportional to the product of the lengths of those intervals.

The length of the intersection of the intervals  $(a,b)$ and $(c,d)$  is
$$ r(a,b,c,d)= \max \{0,
 \min\{ b-a,b-c,d-c,d-a\}\}$$
as can be seen from consideration of all the possible cases. Introducing coordinates  with the origin in the centre of the region between source and detector and $X=(x,y)$ we have $S^{\pm}=(L/2,\pm W_S/2)$ $D^{\pm}=(L/2,\pm W_D/2)$. Equation (\ref{rij}) follows.

\end{document}